\documentclass[aps,prx,preprint,showpacs,superscriptaddress]{revtex4-1} 
\usepackage{graphicx}
\usepackage{epstopdf}

\begin{document}

\title{Taming the degeneration of Bessel beams at anisotropic-isotropic interface: toward 3D control of confined vortical waves}

\author{
Antoine Riaud$^{1,2}$, Jean-Louis Thomas$^{2}$, Michael Baudoin$^{1}$ and Olivier Bou Matar$^{1}$}

\address{$^{1}$Institut d'Electronique, de Micro\'electronique et Nanotechnologie (IEMN), LIA LICS, Universit\'{e} Lille 1 and EC Lille, CNRS UMR 8520, 59652 Villeneuve d'Ascq, France\\
$^{2}$Sorbonne Universit\'{e}s, UPMC Univ Paris 06, CNRS UMR 7588,  Institut des NanoSciences de Paris (INSP), F-75005, Paris, France}



\begin{abstract}
Despite their self-reconstruction properties in heterogeneous media, Bessel beams are known to degenerate when they are refracted from an isotropic to an anisotropic medium. In this paper, we study the converse situation wherein an anisotropic Bessel beam is refracted into an isotropic medium. It is shown that these anisotropic Bessel beams also degenerate, leading to confined vortical waves that may serve as localized particle trap for acoustical tweezers. The linear nature of this degeneration allows the 3D control of this trap position by wavefront correction. Theory is confronted to experiments performed in the field of acoustics. A swirling surface acoustic wave is synthesized at the surface of a piezoelectric crystal by a MEMS integrated system and radiated inside a miniature liquid vessel. The wavefront correction is operated with inverse filter technique. This work opens perspectives for contactless on-chip manipulation devices.
\end{abstract}



\maketitle

\section{Introduction}

\begin{figure}[!h]
\includegraphics[width=80mm]{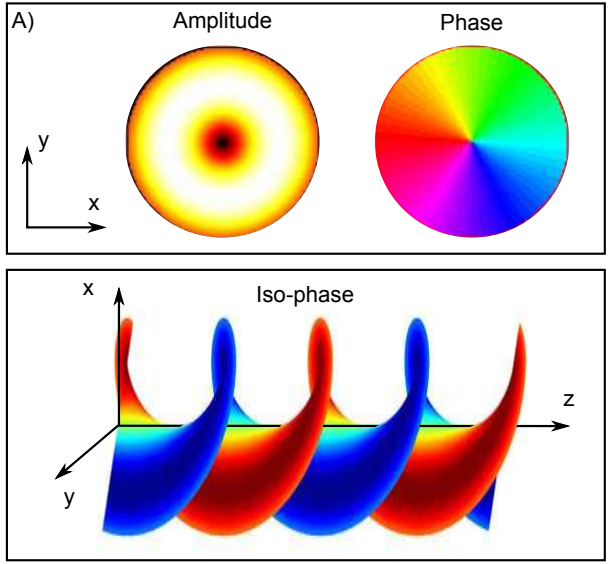}
\caption{Isotropic Bessel beam as defined by equation (\ref{eq: isotropic_bessel}) with topological order $l=1$, axial wave-vector $k_z = 1$ and radial wave-vector $k_r=1$. A) Beam cross-section with complex phase and amplitude. B) Iso-phase surfaces at $l\theta-k_z z = 0$ and $l\theta-k_z z = \pi$ in red and blue respectively. (Color available online)} \label{fig: IsotropicVortex} 
\end{figure}

Bessel beams are helicoidal waves that spiral around a phase singularity (see Fig. \ref{fig: IsotropicVortex})  where their amplitude cancels resulting in concentric rings around a dark core \cite{Allen1992,jasa_hefner_1999}. This singularity can trap objects and is the cornerstone of acoustic and optical tweezers \cite{Marston2006,Baresch2014,Review_tweezers,Trapping_reflecting_particles,Trapping_absorbing_particles,Optical_bottles}. Furthermore, the twisting wavefront transports orbital angular momentum, which can be transmitted to absorbing media and exert a torque. This torque can be used to rotate objects \cite{Allen1992,He1995,MarchianoThomas2003} or generate controlled vortical flows \cite{Riaud2014}. These waves can also be useful for in-depth microscopy since they can self-reconstruct after being damaged by obstacles on the propagation axis \cite{Bessel_beam_reconstr1,Bessel_beam_reconstr2}.  

In spite of this healing ability, numerous reports \cite{isoBessel_refraction1,isoBessel_refraction2,isoBessel_refraction3,isoBessel_refraction4} indicate that refraction between isotropic and anisotropic media is fatal to Bessel beams, as crossing the diopter quickly results in degeneration and fading. In this paper, we show how to predict the fate of anisotropic Bessel beams  \cite{prap_riaud_2015} after refraction into an isotropic medium. Theoretical calculations allows the definition of a coherence length, after which the  vortex dark core shrinks along the propagation axis, resulting in a dark cavity. As long as a dispersion relation can be set for plane waves in both media, our theoretical development allows pre-distorting the wavefront and synthesizing confined vortical waves arbitrarily far away from the interface. 

This effect is illustrated experimentally in the field of acoustics. A confined anisotropic Bessel beam (called a swirling surface acoustic wave)  is synthesized by an integrated MEMS architecture \cite{prap_riaud_2015} and inverse filter technique \cite{Tanter2000,MarchianoThomas2005} at the surface of a piezoelectric anisotropic medium. The refraction of this wave inside a liquid cylinder is measured with a needle hydrophone and the result is compared to our predictions. Then the inverse filter method is used for wavefront pre-distortion and displacement of the acoustical trap (corresponding to the dark cavity) in the x, y and z directions.

Vortex confinement and dark cavity 3D displacement are highly desirable features since they open prospects for dexterous acoustical tweezers. This promising bio-compatible technology \cite{AcousticTweezersReview} could manipulate label-free objects \cite{Baresch2014} with forces up to five order of magnitude above optical tweezers. Compared to individual piezo-ceramic elements \cite{Drinkwater1,Drinkwater2,Drinkwater3,Baresch2014}, interdigitated transducers array \cite{prap_riaud_2015} has major advantages such as simple integration and low cost fabrication with lithography techniques.

\section{From anisotropic Bessel beam refraction to confined vortices}

In this section, we will investigate theoretically the refraction of a Bessel beam from an anisotropic to an isotropic medium. 

For any physical field $f$, refraction of waves between two media is ruled by two conditions: (i) fulfillment of the wave dispersion relation in each media, (ii) continuity conditions at the interface (e.g. continuity of the normal displacement and normal stress in the case of acoustic waves). This continuity condition both induces a spatial matching condition on the complex wave number and determines the transmitted $f_t$ and reflected $f_r$ beams amplitude resulting from an incident wave $f_i$. We use the superposition principle to resolve $f$ into a weighed sum of plane waves. Each of those are defined by a wave-vector associated to a complex magnitude describing the field conserved across the interface. According to acoustic convention, the locus of the wave-vectors will be referred as the slowness surface, and the complex magnitude will be the angular spectrum.

\subsection{Dispersion relation condition}

The plane wave decomposition can be simply obtained through an inverse Fourier transform of $F$, the angular spectrum decomposition of the continuous field $f$. In Cartesian coordinates, we get:
\begin{equation}
f(x,y,z) = \int_{-\infty}^{+\infty}\int_{-\infty}^{+\infty}\int_{-\infty}^{+\infty} F(\kappa_x,\kappa_y,\kappa_z)e^{i \kappa_x x}e^{i \kappa_y y}e^{i \kappa_z z} d\kappa_x d\kappa_y d\kappa_z
\end{equation}
where $F$ is the Fourier Transform of $f$, $(x,y,z)$ are the Cartesian coordinates in the physical space and $(\kappa_x, \kappa_y, \kappa_z)$ the Cartesian coordinates in the reciprocal space. Time dependence is omitted here for the sake of simplicity.

Changing variables to cylindrical coordinates gives: $\kappa_x = \kappa_r \cos(\phi)$,  $\kappa_y = \kappa_r \sin(\phi)$, $x = r\cos(\theta)$ and $y = r\sin(\theta)$, with $(r, \theta, z)$ the cylindrical coordinates in the physical space and $(\kappa_r, \phi, \kappa_z)$ the cylindrical coordinates in the reciprocal space. Straightforward trigonometric algebra yields:

\begin{equation}
f(r,\theta,z) = \int_{-\infty}^{+\infty}\int_{-\pi}^{+\pi}\int_{0}^{+\infty} F(\kappa_r,\phi,\kappa_z)e^{i \kappa_r r \cos(\phi-\theta)}e^{i \kappa_z z} \kappa_r d\kappa_r d\phi d\kappa_z
\end{equation}

For wave fields, the dispersion relation sets a single wave-number $k$ per direction at a given frequency: $F(\kappa_r,\phi,\kappa_z) = h(\phi,\kappa_z)\delta(\kappa_r - k_r(\phi,\kappa_z))$, with $k_r(\phi,\kappa_z) = \sqrt{k^2(\phi,\kappa_z) - \kappa_z^2}$. The resulting integral is a general expression for a wave field in an arbitrary material:
\begin{equation}
f(r,\theta,z) = \int_{-\pi}^{+\pi}\int_{-\infty}^{+\infty} h(\phi,\kappa_z) e^{i k_r(\phi,\kappa_z) r \cos(\phi-\theta)}e^{i \kappa_z z} k_r(\phi,\kappa_z) d\phi d\kappa_z
\label{eq1}
\end{equation}

Now, let $\mathbf{k_i}(\phi,\kappa_z)$ be the wave vector of the incident field in the anisotropic medium and $\mathbf{k_t}(\phi,\kappa_z)$ the transmitted wave-vector. In equation (\ref{eq1}), the angular spectrum $h(\phi,\kappa_z)$ is the only degree of freedom. For diverse reasons, we can simplify further this equation. First, the dispersion relation of the isotropic medium is azimuthal-invariant:
\begin{equation}
f_t(r,\theta,z) = \int_{-\infty}^{+\infty}\int_{-\pi}^{+\pi} h_t(\phi,\kappa_z) e^{i k_{t,r}(\kappa_z) r \cos(\phi - \theta)}k_{t,r}(\kappa_z) d\phi e^{i\kappa_z z}d\kappa_z 
 \label{eq: isotropic_angular spectrum}
\end{equation}  
And second, only the field at the interface is relevant when computing the refraction:
\begin{equation}
f_t(r,\theta,z=0) = \int_{-\infty}^{+\infty}\int_{-\pi}^{+\pi} h_t(\phi,\kappa_z) e^{i k_{t,r}(\kappa_z) r \cos(\phi - \theta)}k_{t,r}(\kappa_z) d\phi d\kappa_z 
 \label{eq: isotropic_angular spectrum_interface}
\end{equation}  

 Similarly to the isotropic medium we only need to evaluate the integral at $z=0$ for the anisotropic medium:

\begin{equation}
f_i(r,\theta,z=0) = \int_{-\infty}^{+\infty}\int_{-\pi}^{+\pi}  h_i(\phi,\kappa_z) e^{i k_{i,r}(\phi,\kappa_z) r \cos(\phi - \theta)}k_{i,r}(\phi,\kappa_z) d\phi d\kappa_z  
 \label{eq: anisotropic angular spectrum}
\end{equation}  

\begin{figure}[htbp]
\includegraphics[width=80mm]{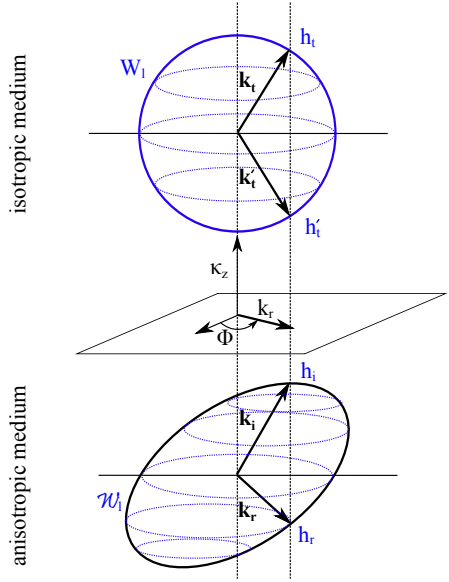}
\caption{Angular spectrum representation of the refraction of a plane wave from an anisotropic to an isotropic medium. The incident beam represented by its angular spectrum $h_i(\phi,\kappa_z)$ and wave-vector $\mathbf{k}_i$ splits into a reflected field $h_r(\phi,\kappa_z)$ and a transmitted field $h_t(\phi,\kappa_z)$. The dispersion relation condition appears graphically by forcing all wave-vectors to lie on the slowness surface of their respective propagation media. The spatial matching boundary condition forces the wave-vectors to share the same propagation direction $\phi$ and  radial component  $k_r=\mathbf{k}_{i,r}=\mathbf{k}_{t,r}=\mathbf{k}_{r,r}$.
The transmitted (reflected) field then happens as the intersection between the isotropic (anisotropic) media slowness surface and the set of all wave-vectors with radial component $k_r$. The wave-vector $\mathbf{k}'_t$ is oriented towards the anisotropic medium and cannot exist. Isotropic and anisotropic Bessel waves $W_l$ and $\mathcal{W}_l$ (dotted blue circles) are the intersection between the slowness surfaces and planes normal to the propagation axis. (Color available online)} \label{fig: angular spectrum perspective on transmission}  \end{figure}

The two integrals (\ref{eq: isotropic_angular spectrum_interface}) and (\ref{eq: anisotropic angular spectrum}) represent the ensemble of waves that could exist in each medium depending on the values of $h_i$ and $h_t$. They are represented in the reciprocal space in figure \ref{fig: angular spectrum perspective on transmission}: $h_i(\phi,\kappa_z)$ and $h_t(\phi,\kappa_z)$ are bi-variable functions of complex value on a spherical manifold for $h_t$ and an arbitrary surface for $h_i$. In order to match the conserved field (e.g. normal displacement and normal stress) in both media at a given $z=0$, we have to balance equations (\ref{eq: isotropic_angular spectrum_interface}) and (\ref{eq: anisotropic angular spectrum}). This in-plane spatial matching also implies a spectral one in the reciprocal space. This means that the projection of both fields on the plane normal to the propagation axis $\mathbf{\kappa}_z$ must match. The graphical solution then appears as the intersection of both media slowness surfaces for all possible $\kappa_z$.

\subsection{Spatial matching condition}

Besides graphical construction, the refraction of the wave field can be computed with the angular spectrum. This framework allows considering each plane wave independently. The continuity condition of the physical field $f$ across the interface plane reads $f_i + f_r = f_t$ over all the spatial extent of the experiment. Assuming it is much larger than the wavelength, the same condition applies on the angular spectra $h_i + h_r = h_t$ for any particular $\kappa_r$ and $\phi$ while keeping $\kappa_z$ a free parameter. We will refer to $\kappa_{i,z}$, $\kappa_{r,z}$, $\kappa_{t,z}$ for the incident, reflected and transmitted waves. Using the transmission coefficient $t(\phi,\kappa_z)$, we can express $h_t(\phi,\kappa_{t,z}) = t(\phi,\kappa_{i,z})h_i(\phi,\kappa_{i,z})\delta(k_{i,r}(\phi,\kappa_{i,z}) - k_{t,r}(\phi,\kappa_{t,z}))$, with $\delta$ the Dirac operator describing the intersection between the slowness surfaces of the isotropic medium and the anisotropic one. Hence, integral (\ref{eq: isotropic_angular spectrum}) reduces to:

\begin{equation}
f_t(r,\theta,z) = \int_{-\infty}^{+\infty}\int_{-\pi}^{+\pi}  t(\phi,\kappa_{i,z}) h_i(\phi,\kappa_{i,z}) e^{i k_{i,r}(\phi,\kappa_{i,z}) r \cos(\phi - \theta) + i k_z(\phi,\kappa_{i,z})z}k_{i,r}(\phi,\kappa_{i,z}) d\phi d\kappa_{i,z} 
\label{eq: refracted angular spectrum total}
\end{equation}

The Dirac operator links $\kappa_{i,z}$, $\kappa_{r,z}$, $\kappa_{t,z}$ and enforces a new dispersion relation $k_z(\phi,\kappa_{i,z}) = \kappa_{t,z}(\phi,\kappa_{i,z}) = \pm\sqrt{k_t^2 - k_{i,r}^2(\phi,\kappa_{i,z})}$. \footnote{In some media, this equation might admit no real solution, in which case no wave would transmit from a media to another, but evanescent guided waves (e.g. surface plasmons, Scholte and Stoneley waves) would still appear.} The locus $(k_{i,r}(\phi,\kappa_{i,z}),k_z(\phi,\kappa_{i,z}))$ describes the slowness surface of any wave that could possibly be transmitted from the anisotropic to the isotropic medium. This important result implies that any cross-section of the wave field orthogonal to $\mathbf{z}$ must fulfill the source-medium dispersion relation. Reversely, given an arbitrary reference altitude $z_0$ we can synthesize any wave that could be formed on the anisotropic medium surface simply by choosing $h_i(\phi,\kappa_{i,z}) = h_i'(\phi,\kappa_{i,z})e^{-i k_z(\phi,\kappa_{i,z})z_0}$. 

\subsection{Field refracted by an anisotropic Bessel beam}
\label{ss:abb}

In the following, we assume the beam in the anisotropic medium to be either a surface or a bulk wave. In the first case, the wave travels along the interface, so $h_i \neq 0$ only for $\kappa_{i,z} = 0$. In the second case, we will only consider waves that keep their coherence while propagating along z-axis in the anisotropic medium. For Bessel beams, this latter condition amounts to saying that we are able to define some anisotropic Bessel beams that remain coherent while propagating along the z-axis. Following this assumption, the incident beam angular spectrum must lie on the meridian $\kappa_z = k_{i,z}$. Hence, in both cases  $h_i(\phi,\kappa_{i,z}) = h_i(\phi)\delta(\kappa_{i,z}-k_{i,z})$. Equation (\ref{eq: refracted angular spectrum total}) becomes:

\begin{equation}
f_t(r,\theta,z) = \int_{-\pi}^{+\pi} t(\phi) h_i(\phi) e^{i k_r(\phi) r \cos(\phi - \theta) + i k_z(\phi)z}k_r(\phi) d\phi  
\label{eq: refracted angular spectrum}
\end{equation}
Where we used the short-hand notations  $t(\phi) = t(\phi,k_{i,z})$, $k_r(\phi) = k_{i,r}(\phi,k_{i,z})$ and $k_z(\phi) = k_z(\phi,k_{i,z})$.
In our previous work \cite{prap_riaud_2015}, we introduced a definition of  anisotropic Bessel beams  $\mathcal{W}_l = \mathcal{W}_l^0(r,\theta)e^{ik_z z}$, which are coherent solutions in cylindrical coordinates of the wave equations in anisotropic media, with $\mathcal{W}_l^0$ the so-called anisotropic swirling wave: 
\begin{equation}
\mathcal{W}_l^0(r,\theta) = \frac{1}{2\pi i ^l}\int_{-\pi}^{\pi} e^{il\phi + i k_{r}(\phi)r\cos(\phi-\theta)}d\phi
\label{eq: anisotropic swirling SAW}
\end{equation}
where $l$ is the topological order of the anisotropic Bessel beam.

In isotropic media, the integral definition of the Bessel function with a straightforward change of variable $\eta = \pi/2 + \phi-\theta $ yields $\mathcal{W}_l^0 = J_l e^{i l \theta}$, leading to the more common isotropic Bessel beam $W_l$:
\begin{equation}
W_l = J_l(k_r r) e^{i l \theta +i k_z z}
\label{eq: isotropic_bessel}
\end{equation}
Isotropic and anisotropic Bessel beams are defined by their angular spectrum $h = e^{il\phi}/(2\pi i^l k_r(\phi,\kappa_z))$. In order to create an anisotropic swirling wave $\Psi_l$ of topological order $l$ at an arbitrary altitude $z_0$, we set $h_i(\phi) = e^{i l \phi - i k_z z_0}/(2\pi i^l k_r(\phi))$. Thus, equation (\ref{eq: refracted angular spectrum}) yields:

\begin{equation}
\Psi_l(r,\theta,z) = \frac{1}{2\pi i^l}\int_{-\pi}^{\pi} t(\phi) e^{il\phi+i k_{r}(\phi) r \cos(\phi - \theta)+i k_z(\phi) (z-z0)} d\phi 
\label{eq: diffracted Bessel beam}
\end{equation}

It is easily inferred from equation (\ref{eq: diffracted Bessel beam}) that the small azimuthal variations of $k_z$ will be dramatically magnified by increasing values of $|z-z_0|$, resulting in beam degeneration.

\section{Vortex strengthening, confinement and hole filling}

For moderately anisotropic media, we can get further insight on this degeneration by introducing a simplified slowness surface: $k_{i,r} = k_i^0(1+\epsilon \sin(2\phi))$, with $\epsilon << 1$. The computation, detailed in the appendix yields:

\begin{equation}
\Psi_l(r,\theta,z) \simeq  t^0 e^{i k^0_z z} \mathcal{W}_{l}^0(r,\theta) J_0(\frac{{k_i^0}^2\epsilon z}{k_z^0}) +
  t^0 e^{i k^0_z z} \sum\limits_{n = 1}^{+\infty} J_n(\frac{{k_i^0}^2\epsilon z}{k_z^0})\left[ \mathcal{W}_{l+2n}^0(r,\theta)+(-1)^n\mathcal{W}_{l-2n}^0(r,\theta) \right]   
\label{eq: scattered bessel}
\end{equation}
With ${k^0_z}^2 = {k_t}^2 - {k^0_i}^2$ and $t^0$ the average transmission coefficient over all directions. Equation (\ref{eq: scattered bessel}) is made of two terms: a fundamental beam $\mathcal{W}_{l}^0$ plus a series of degenerated ones $\mathcal{W}_{l\pm 2n}^0$. All these beams are only functions of $r,\theta$, but have a $z$-dependent factor $J_n(\frac{{k_i^0}^2\epsilon z}{k_z^0})$. We can note that the same coefficients also appears in frequency modulation signal modeling. Indeed, the current phenomenon is highly similar to spatial wavelength modulation since the wavelength in anisotropic media is generally direction $\phi$ dependent. For $z$ close to zero, the zero-order Bessel function predominates, such that the beam is similar to the fundamental. For $z$ large enough, the $J_0(\frac{{k_i^0}^2\epsilon z}{k_z^0})$ function cancels, so the fundamental mode eventually vanishes. In the meantime, the other harmonics and especially the first mode grow in amplitude, and reach the fundamental for the first time at $z = \Lambda$ with $\Lambda$ solution of $J_0(\frac{{k_i^0}^2\epsilon \Lambda}{k_z^0}) = J_1(\frac{{k_i^0}^2\epsilon \Lambda}{k_z^0})$ being the coherence length:
\begin{equation}
\Lambda \simeq 1.43\frac{k_z^0}{{k_i^0}^2\epsilon} 
\end{equation}

We can also infer that for even order beams, the degeneration given by equation (\ref{eq: scattered bessel}) will happen through vortex hole filling. With increasing values of $n$, the terms $\mathcal{W}_{l - 2n}^0$ and $\mathcal{W}_{l + 2n}^0$ are magnified by their coefficients $J_n(\frac{{k_i^0}^2\epsilon z}{k_z^0})$. Bessel beams of decreasing order have smaller dark cores, so for growing $n$, $\mathcal{W}_{l - 2n}^0$ is magnified; it becomes the dominant mode and the vortex dark core shrinks. Eventually, provided the vortex is of even topological charge $l$, a large enough value of $n = l/2$ will imprint a $\mathcal{W}_0^0$ wave as local maximum of amplitude on the vortex axis, filling the Bessel dark core. This happens at the first local extremum of $J_n(\frac{{k_i^0}^2\epsilon z}{k_z^0})$. The solution of $J_n'(x) = 0$ is often written $j'_{n,1}$ and is given by Olver \cite{Olver1951} formula: $j'_{l/2,1} = l(1+0.809\times (l/2)^{-2/3} + O(l^{-4/3}))/2$. If we solve $\frac{{k_i^0}^2\epsilon \Delta_{\mathtt{hole}}}{k_z^0} = j'_{l/2,1}$, we have:
\begin{equation}
\Delta_{\mathtt{hole}} \simeq \frac{lk^0_z\left(1+1.284 \times l^{-2/3}\right)}{2\epsilon {k^0_i}^2} 
\label{eq: depth of hole}
\end{equation}

According to this equation, the depth of hole of isotropic vortices ($\epsilon=0$) would be infinite. 

To summarize, the following properties are expected when an anisotropic Bessel beam is diffracted by an isotropic medium:  (i) the vortex can be formed far away from the substrate (ii) the beam degenerates after a distance $\Lambda$ which depends on the strength of the anisotropy and (iii) even order vortices exhibit hole filling.

As mentioned in the introduction, these effects are very favorable to acoustical trapping and manipulation. In the following, we test experimentally our predictions and prove that confined Bessel beams can be generated from surface acoustic wave integrated devices, possibly solving the trade-off between cost and selectivity. 

\section{Experimental setup}

\begin{figure}[htbp]
\includegraphics[width=80mm]{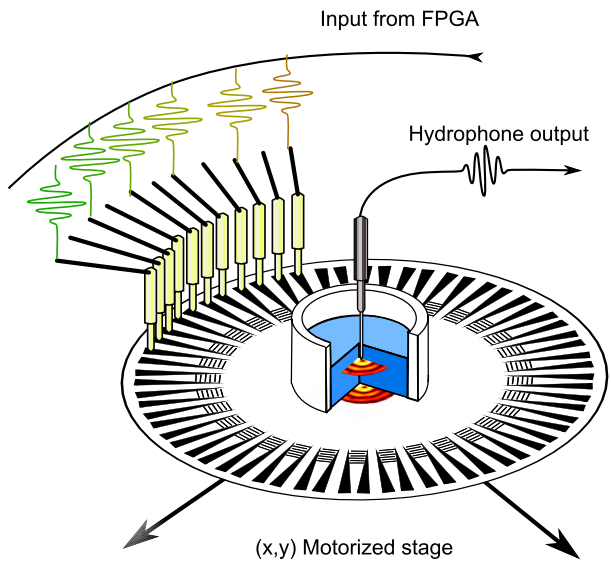}
\caption{Experimental setup. (Color available online)} 
\label{fig: experimental_setup}  
\end{figure}

The experimental setup, shown in figure (\ref{fig: experimental_setup}), is similar to the one previously used by the authors \cite{prap_riaud_2015} to synthesize swirling SAW. Surface acoustic waves are generated at the surface of a 8 inches 1 mm thick X-cut lithium niobate piezo-electric wafer by an array of 32 interdigitated transducers powered by a field programmable gate array (FPGA - Lecoeur Electronics), whose input signal is designed using two-dimensional inverse filter \cite{Tanter2000}.  A small PVC cylinder is sealed at the center of the wafer (inner diameter = 14.1 mm, height = 8 mm, seal in silicon) and filled with water to study the refraction of the anisotropic swirling wave into an isotropic medium. The acoustic field is recorded in water using a 75 $\mu$m diameter hydrophone (Precision Acoustics LTD) whose position is controlled by a vernier (vertical displacement) and 2 micro-stepper motors (horizontal motion). 

Each experiment was performed following the same protocol. First, we calibrate our 32 transducers array by scanning a 2 mm diameter virtual disc located 2 mm above the substrate. Then, we compute the optimal input with the inverse filter method to get the desired wave field. Neglecting diffraction, there is a unique relation between $h_i(\phi)$ and $h_t(\phi)$. So setting the field in a single plane gives $h_i(\phi)$, which amounts to choosing $h_t(\phi)$ in the whole fluid cavity. After programming the FPGA, we measure the emitted acoustic field in the regions of interest using the needle hydrophone. 

\subsection{Confined acoustical vortex structure}

\begin{figure*}[htbp]
\includegraphics[width=140mm]{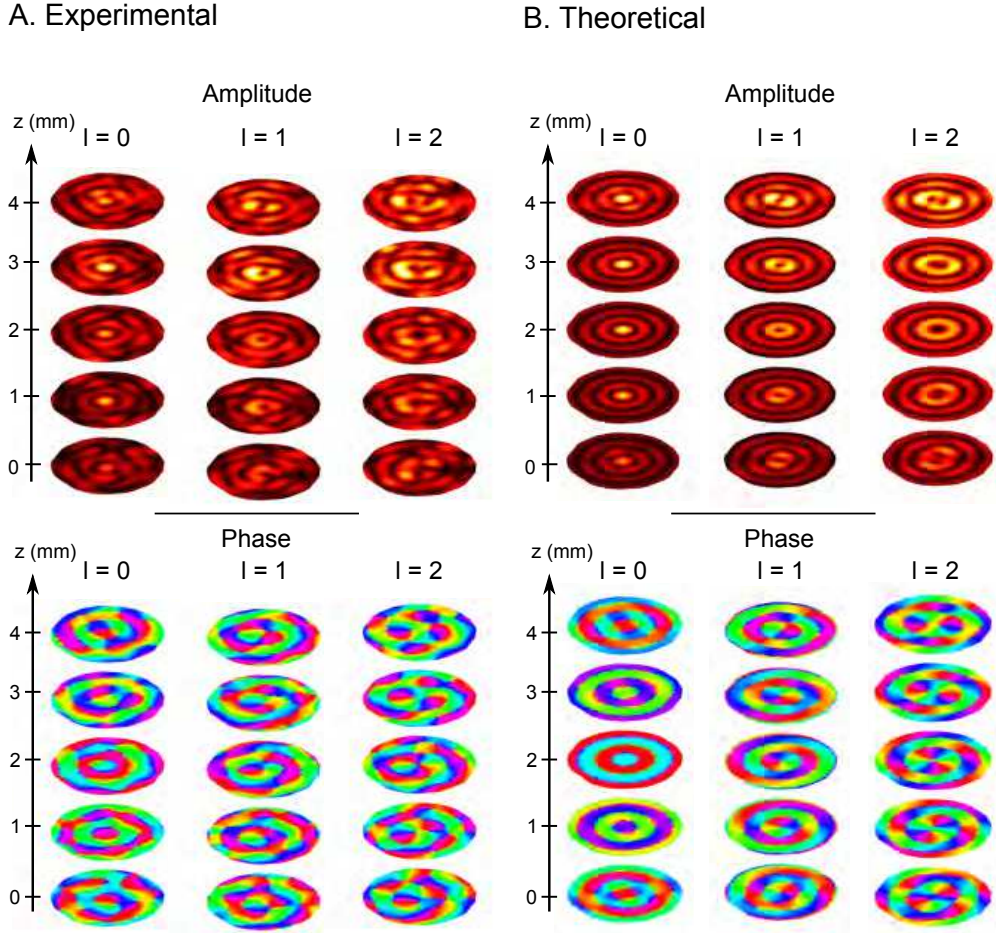}
\caption{Anisotropic Bessel beams propagating in an isotropic fluid. The reference plane is located at $z_0=2$ mm. On this plane, the first bright ring radii scale linearly to the topological order and are respectively $100$ and $200$ $\mu$m for $l=1$ and $l=2$. The beams are expected to change on a characteristic distance $\Lambda \simeq 2.2$ mm and have their dark core filled after $\Delta_{hole} \simeq 1.8$ and $2.8$ mm for the first and second order vortices respectively. The plotted discs have a 1 mm diameter. \textbf{A.} Experimental synthesis of confined acoustical vortices. Topological orders range from 0 to 2. Max pressure amplitudes are respectively 0.35, 0.24, and 0.26 MPa. \textbf{B.} Corresponding theoretical waves computed from equation (\ref{eq: diffracted Bessel beam}). (Color available online)
\label{fig: various_orders} } 
\end{figure*}

We synthesized vortices of order $l=$0, 1 and 2 at $z_0 = 2$ mm above the substrate. Results are shown in figure \ref{fig: various_orders}.A.

In figure \ref{fig: various_orders}.B, we display the numerical integration of equation (\ref{eq: diffracted Bessel beam}) of the same vortices. The experimental pressure fields are qualitatively and quantitatively similar to the predictions for all three topological orders. The experimental field appears slightly blurrier than the theoretical one, most likely due to diffraction as the wave meets the three-phase interfaces (air/silicone/substrate and silicone/water/substrate) and possibly to the discrete number of transducers used in our experiments that limits the synthesis of the prescribed wave field. The first bright ring radii scale linearly to the topological order and are respectively $100$ and $200$ $\mu$m for $l=1,2$. Despite this lower signal quality, the expected effects still appear quite clearly, with the strengthening of the beams, the degeneracy before and after the reference plane $z_0 = 2$ mm and the hole filling for the second order vortex. For lithium niobate in water at 12 MHz, $\Lambda \simeq 2.2$ mm.

Taking the $l=0$ case, we notice that the beam strengthens as it gets further away from the substrate. This surprising effect can be explained simply by tracing the origin of the beam. When a surface acoustic wave penetrates in the liquid, it starts radiating some power $\mathcal{P}_0$ along the Rayleigh angle $\theta_R \simeq 25^o$. Acoustic waves dissipation in water being very weak, these bulk waves reach the vortex axis at $ z = D/2\tan(\theta_R)$ almost unattenuated. On the opposite, the SAW gradually loses power as it propagates under the liquid layer, hence, the wave combining at $z=0$ will have a power $\mathcal{P}_0e^{-\alpha \omega D/(4\pi)}$, with $\alpha \simeq 0.20$ dB/MHz.mm \cite{CampbellJones}. Following geometrical arguments, the characteristic gain length is $l_{\mathtt{gain}} = \omega\alpha\tan(\theta_R)/(2\pi) \simeq 1.1$ dB/mm at 12 MHz. Interestingly, some authors proposed that gain media could lead to tractor beam \cite{Gain_media_tractor_beam}.

\begin{figure}[htbp]
\includegraphics[width=80mm]{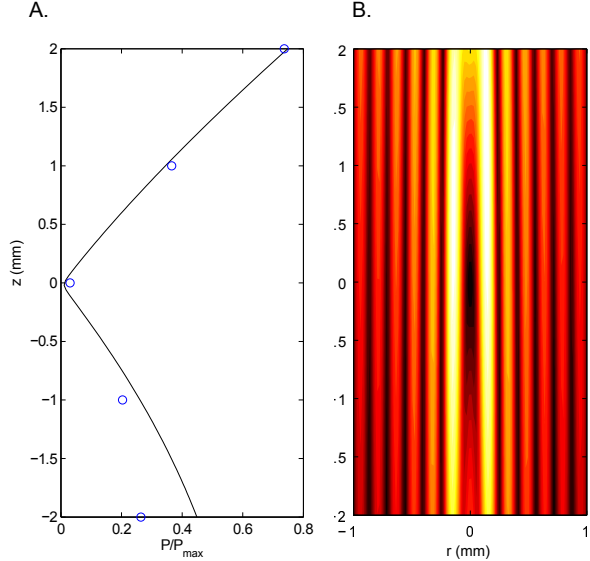}
\caption{Anisotropic second order Bessel beam propagating in an isotropic fluid. The reference plane is located at $z_0=0$ mm. \textbf{A.} Pressure complex amplitude on the vortex axis for various altitudes $z$ normalized by the maximum pressure amplitude of the first bright ring at $z=0$. Blue circles represent experimental data points and the black solid line the corresponding theoretical amplitude. \textbf{B.} Cross-section of the complex pressure amplitude deduced from equation (\ref{eq: diffracted Bessel beam}). (Color available online)
\label{fig: cross_section_vortex} } 
\end{figure}

The two other effects are more visible on higher order confined vortices. Looking at the phase of the $l=2$ vortex, we notice that the phase singularity quickly degenerates for increasing $|z-z_0|$ values and its amplitude bright ring appears squeezed. Similarly, the ring of the first order vortex splits at $|z-z_0| > 1$ mm. For $l=2$, this beam degeneration is associated to hole filling, as for increasing $|z-z_0|$ the beam intensity on the axis grows up, filling the vortex hole. This is more easily visualized by taking the cross-section of the degenerated vortices, shown in figure \ref{fig: cross_section_vortex}.

The confinement of the vortices suggests to attribute them $(x,y,z)$ coordinates. In the following, we prove experimentally the 3D mobility of these confined acoustical vortices, confirming their potential use for acoustic tweezers. 

\subsection{x-y mobility}

\begin{figure*}[htbp]
\includegraphics[width=120mm]{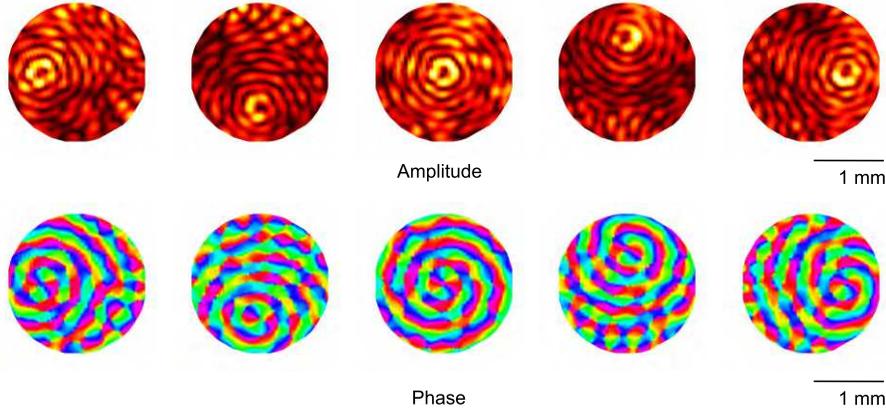}
\caption{Synthesis of 2nd order confined acoustical vortices at different positions 2 mm above the substrate. Maximum pressure amplitude of each vortex is 0.18 MPa. (Color available online)} 
\label{fig: mobility_xy}  
\end{figure*}

As a rule of thumb, larger magnification results in smaller depth of field, so labs on a chip are essentially 2D devices. The $(x,y)$ mobility in the plane parallel to the substrate is therefore critical. In the present experiment, we used the same calibration to generate vortices at five different $(x,y)$ positions, with the acoustic pressure amplitude and phase displayed in figure \ref{fig: mobility_xy}.  

The wave field faithfully reproduces the required displacement over 1 mm (about 15 wavelengths). This confirms the applicability of confined vortices for 2D manipulation of micro-objects.

\subsection{z mobility}

\begin{figure}[htbp]
\includegraphics[width=80mm]{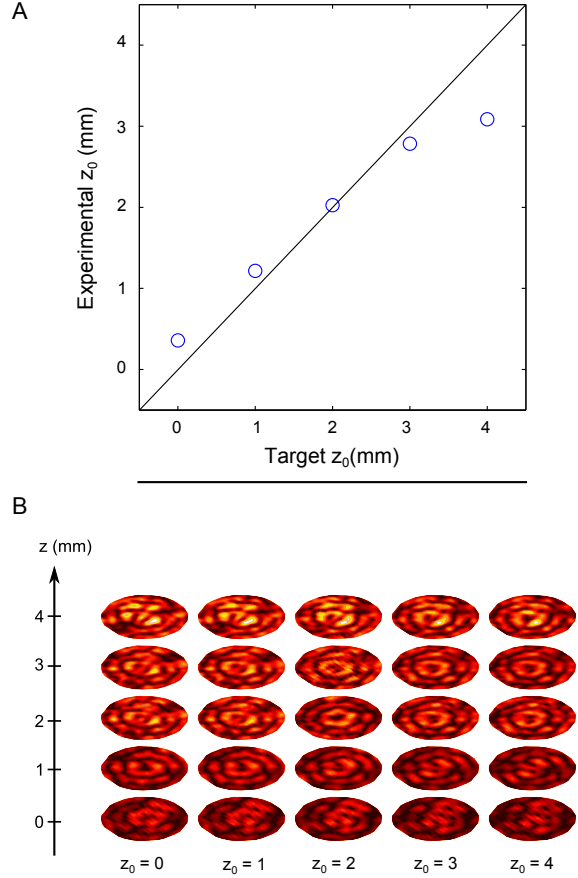}
\caption{Synthesis of 2nd order confined acoustical vortices at different altitudes $z_0$. A) Maximum correlation coefficient. B) Experimental fields.  Max pressure amplitudes are respectively 0.22, 0.24, 0.26 0.30 and 0.33 MPa. (Color available online)} 
\label{fig: mobility_z2}  
\end{figure}

One of the largest expectation on confined vortices is the vertical manipulation of objects. This additional degree of freedom is  essential for the manipulation of biological objects like cells or bacteria that can strongly interact with walls. This interaction can be desirable to achieve cell sorting according to the cells adhesion properties \cite{bussonniere2014cell} or detrimental in experiments where walls effects are not the purpose of the study. The z-mobility is also useful when studying bubbles or contrast agents properties which are systematically drived close to the walls by buoyancy. Finally, it might also allow manipulating multiple stacked labs on chips. In this experiment, we kept using the calibration obtained from $z = 2$ mm, but synthesized cross-section of vortices with different reference plane $z_0$ ranging from 0 to 4 mm above the substrate. Results are shown in figure \ref{fig: mobility_z2}.  

After the experiment, the data were processed to extract the position of the phase singularity. The analysis starts by computing the cross-correlation between a confined vortex from equation (\ref{eq: diffracted Bessel beam}) and the experimental measurements. The maximum of the correlation is labeled as the experimental $z_0$, whereas the one used to design the vortices is tagged target $z_0$. The general good agreement between target and experimental values shown in figure \ref{fig: mobility_z2}.A confirms the vertical mobility of confined vortices. The $z-$mobility is however bounded by the spatial extent of the calibration area and the Rayleigh radiation angle. Indeed, waves coming from beyond the edge of the calibration area are ill-defined and reach the central axis $z=D_{\mathtt{cal}}/2\tan(\theta_R) \simeq 2.1$ mm away from the calibration plane, bringing some error to the wave propagation.

Visual comparison between vortices of different $z_0$ in figure \ref{fig: mobility_z2}.B agrees well with the correlation trend. Here, each column represents a wave with a given target $z_0$, whereas each line is a different measurement plane $z$. If the vertical motion was ideal, one could expect the waveforms to be invariant when following the diagonals $z-z_0 = cst$. For instance, taking the central diagonal ($z = z_0$) we can see that the dark core appears on all scans, whereas on the diagonal $z-z_0 = \pm 2$ mm the hole filling is almost always at the same development stage.  

In this experimental section, we successfully synthesized confined vortices  at a controllable $(x,y,z)$ position. The beams degenerate in excellent agreement with our theory and exhibit the researched effect of hole filling. This opens prospects for affordable and dexterous contactless manipulation devices, which will be our future objective. Beyond acoustics, we believe depth of hole and vortex confinement could find numerous applications in other physical fields involving anisotropic wave propagation. 

\section{Conclusion}

In this work, we address the general problem of the refraction of an anisotropic Bessel beam into an isotropic medium. After getting an explicit expression of the refracted field, we derive a coherence length beyond which the beam starts to degenerate. Furthermore, we show both theoretically and experimentally that it is possible to anticipate this degeneration and form phase singularity arbitrarily far away from the interface. For even topological orders, the resulting beam exhibits hole filling. This progressive filling of the vortical wave dark core opens prospects for contactless manipulation in both optics and acoustics. In optics, similar approach could also allow in depth microscopy in anisotropic media.

\begin{acknowledgments}
This work was supported by ANR project ANR-12-BS09-0021-01.
\end{acknowledgments}

\appendix*

\section{Simplification of the refracted beam for moderately anisotropic materials}

In the following, and without loss of generality, we choose $z_0 = 0$. For weak anisotropy, we assume the slowness contour of the solid to be $k_{i,r}(\phi) = k^0_i(1+\epsilon \sin(2\phi))$. Using the dispersion relation of the isotropic medium $k_z^2 + k_{i,r}^2 = {k_t}^2 = \omega^2/c_i^2$, we easily get $k_z(\phi) \simeq k_z^0(1-(k_i^0/k^0_z)^2\epsilon\sin(2\phi))$, with ${k_z^0}^2 = {k_t}^2 - {k_i^0}^2$. The exponential term $e^{-i{k_i^0}^2 z\epsilon\sin(2\phi)/k_z^0}$ is then developed according to the Jacobi-Anger expansion, it yields:

\begin{equation}
\Psi_l(r,\theta,z) \simeq  \frac{e^{i k^0_z z}}{2\pi i^l}\int_{-\pi}^{\pi} t(\phi) e^{il\phi+i k_i(\phi) r \cos(\phi - \theta)}\sum\limits_{n = -\infty}^{+\infty} (-1)^n e^{2 i n\phi}J_n(\frac{{k_i^0}^2\epsilon z}{k_z^0}) d\phi 
\end{equation}
After grouping the terms, we get:
\begin{equation}
\Psi_l(r,\theta,z) \simeq  e^{i k^0_z z} \mathcal{W'}_{l}^0(r,\theta) J_0(\frac{{k_i^0}^2\epsilon z}{k_z^0}) +
  e^{i k^0_z z} \sum\limits_{n = 1}^{+\infty} J_n(\frac{{k_i^0}^2\epsilon z}{k_z^0})\left[ \mathcal{W'}_{l+2n}^0(r,\theta)+(-1)^n\mathcal{W'}_{l-2n}^0(r,\theta) \right]   
\end{equation}
With $\mathcal{W'}_{l}^0$ the refracted Bessel beam distorted by the anisotropic transmission coefficient:
\begin{equation}
\mathcal{W'}_l^0(r,\theta) = \frac{1}{2\pi i ^l}\int_{-\pi}^{\pi} t(\phi) e^{il\phi + i k_{i,r}(\phi)r\cos(\phi-\theta)}d\phi
\label{eq: refracted swirling SAW}
\end{equation}
In practice, we can take $\mathcal{W'}_{l}^0 \simeq t^0 \mathcal{W}_{l}^0$ for two reasons: (i) errors in complex magnitude on the angular spectrum keep constant during the wave propagation, whereas errors in phase grow exponentially and (ii) most materials are weakly anisotropic such that $t^0$ the average value of $t(\phi)$ provides a good approximation of the overall transmission coefficient. This yields equation (\ref{eq: scattered bessel}).

\bibliography{AnisotropicVortexBibliography2}





\label{lastpage}
\end{document}